\documentclass[twocolumn]{aastex}
\usepackage{amsmath,esint}
\usepackage{url,aasmacros}
\usepackage{natbib}
\usepackage{lineno}
\usepackage{soul, CJK}
\usepackage{multirow}

\usepackage{bm}
\usepackage{dirtytalk}
\usepackage{xspace}
\usepackage{color}
\usepackage{enumitem}

\def\mr{\mathrm}

\def\t{\widetilde}
\def\mc{\mathcal}

\newcommand{\lara}[1]{{\langle#1\rangle}}
\renewcommand*{\thefootnote}{\fnsymbol{footnote}}
\newcommand{\frb}{FRB 180916.J0158+65}

\usepackage[utf8]{inputenc}

\shorttitle{ZTF optical limits on \frb{}}
\shortauthors{Andreoni et al.}

\begin{document}

\title{	
Zwicky Transient Facility constraints on the optical emission from the nearby repeating FRB~180916.J0158+65}

\author[0000-0002-8977-1498]{Igor~Andreoni$^*$}
\email{$^*$andreoni@caltech.edu}
\affiliation{Division of Physics, Mathematics and Astronomy, California Institute of Technology, Pasadena, CA 91125, USA}

\author{Wenbin Lu}
\affiliation{Division of Physics, Mathematics and Astronomy, California Institute of Technology, Pasadena, CA 91125, USA}

\author{Roger M. Smith}
\affiliation{Caltech Optical Observatories, California Institute of Technology, Pasadena, CA 91125, USA}

\author[0000-0002-8532-9395]{Frank J. Masci}
\affiliation{IPAC, California Institute of Technology, 1200 E. California Blvd, Pasadena, CA 91125, USA}

\author[0000-0001-8018-5348]{Eric C. Bellm}
\affiliation{DIRAC Institute, Department of Astronomy, University of Washington, 3910 15th Avenue NE, Seattle, WA 98195, USA}

\author{Matthew J. Graham}
\affiliation{Division of Physics, Mathematics and Astronomy, California Institute of Technology, Pasadena, CA 91125, USA}

\author{David L. Kaplan}
\affiliation{Center for Gravitation, Cosmology, and Astrophysics, Department of Physics, University of Wisconsin-Milwaukee, P.O. Box 413,
Milwaukee, WI 53201, USA}

\author{Mansi M. Kasliwal}
\affiliation{Division of Physics, Mathematics and Astronomy, California Institute of Technology, Pasadena, CA 91125, USA}

\author{Stephen Kaye}
\affiliation{Caltech Optical Observatories, California Institute of Technology, Pasadena, CA 91125, USA}

\author[0000-0002-6540-1484]{Thomas Kupfer}
\affiliation{Kavli Institute for Theoretical Physics, University of California, Santa Barbara, CA 93106, USA}

\author[0000-0003-2451-5482]{Russ R. Laher}
\affiliation{IPAC, California Institute of Technology, 1200 E. California  Blvd, Pasadena, CA 91125, USA}

\author{Ashish A. Mahabal}
\affiliation{Division of Physics, Mathematics and Astronomy, California Institute of Technology, Pasadena, CA 91125, USA}

\author{Jakob Nordin}
\affiliation{Institute of Physics, Humboldt-Universit{\"a}t zu Berlin, Newtonstr. 15, 12489 Berlin, Germany}

\author{Michael Porter}
\affiliation{Caltech Optical Observatories, California Institute of Technology, Pasadena, CA 91125, USA}

\author{Thomas A. Prince}
\affiliation{Division of Physics, Mathematics and Astronomy, California Institute of Technology, Pasadena, CA 91125, USA}

\author{Dan Reiley}
\affiliation{Caltech Optical Observatories, California Institute of Technology, Pasadena, CA 91125, USA}

\author{Reed Riddle}
\affiliation{Caltech Optical Observatories, California Institute of Technology, Pasadena, CA 91125, USA}

\author{Joannes Van Roestel}
\affiliation{Division of Physics, Mathematics and Astronomy, California Institute of Technology, Pasadena, CA 91125, USA}

\author[0000-0001-6747-8509]{Yuhan Yao}
\affiliation{Division of Physics, Mathematics and Astronomy, California Institute of Technology, Pasadena, CA 91125, USA}

\begin{abstract}
The discovery rate of fast radio bursts (FRBs) is increasing dramatically thanks to new radio facilities. Meanwhile, wide-field instruments such as the 47\,deg$^2$ Zwicky Transient Facility (ZTF) survey the optical sky to study transient and variable sources. We present serendipitous ZTF observations of the CHIME repeating source \frb{}, that was localized to a spiral galaxy 149\,Mpc away and is the first FRB suggesting periodic modulation in its activity. While 147 ZTF exposures corresponded to expected high-activity periods of this FRB, no single ZTF exposure was at the same time as a CHIME detection. No $>3\sigma$ optical source was found at the FRB location in 683 ZTF exposures, totalling 5.69 hours of integration time.
We combined ZTF upper limits and expected repetitions from \frb{} in a statistical framework using a Weibull distribution, agnostic of periodic modulation priors. The analysis yielded a constraint on the ratio between the optical and radio fluences of $\eta \lesssim 200$, corresponding to an optical energy $E_{\rm opt} \lesssim 3 \times 10^{46}$\,erg for a fiducial 10\,Jy\,ms FRB (90\% confidence).
A deeper (but less statistically robust) constraint of $\eta \lesssim 3$ can be placed assuming a rate of $r(>5\,\rm Jy\, ms)=1\rm\,hr^{-1}$ and $1.2\pm 1.1$ FRB occurring during exposures taken in high-activity windows.
 The constraint can be improved with shorter per-image exposures and longer integration time, or observing FRBs at higher Galactic latitudes. This work demonstrated how current surveys can statistically constrain multi-wavelength counterparts to FRBs even without deliberately scheduled simultaneous radio observation. 

\end{abstract}

\section{Introduction}
\label{sec: intro}

Fast Radio Bursts (FRBs) are cosmological millisecond-duration radio flashes that are now discovered routinely by facilities such as the Parkes telescope \citep[e.g.,][]{Bhandari2018a}, the Canadian Hydrogen Intensity Mapping Experiment \citep[CHIME;][]{CHIME2018system}, the 
updated Molonglo Observatory Synthesis Telescope \citep[UTMOST; e.g.,][]{Farah2018}, the Australian Square Kilometre Array Pathfinder \citep[e.g.,][]{Shannon2018Nat}, and the Deep Synoptic Array \citep[DSA;][]{Kocz2019DSA, Ravi2019a}. Several FRBs were found to repeat \citep{Spitler2016, Kumar2019, CHIME2019rep2, CHIME2019rep, Chime2020rep}, which suggests that a fraction \citep[if not all, see][]{Ravi2019b} of FRB progenitors are not disrupted by the outburst.

Searches for optical or high-energy counterparts were conducted during standard ``triggered" follow-up observations \citep[e.g.,][]{Petroff2015, Shannon2017, Bhandari2018a} as well as during simultaneous observations with wide-field telescopes \citep[e.g.,][]{DeLaunay2016, Tingay2019, Martone2019} or targeting repeating FRBs \citep[e.g.,][]{Scholz2016, Scholz2017, Hardy2017, MAGIC2018}. A marginal gamma-ray candidate was found by \cite{DeLaunay2016} possibly associated with FRB~131104, but a robust transient counterpart to a FRB is yet to be discovered. The recent detection of a bright $>1.5$\,MJy\,ms radio burst probably associated with the Galactic soft gamma-ray repeater SGR~1935+2154 \citep{Bochenek2020ATel} may offer an important piece to solve the FRB puzzle. 

A particularly interesting repeating source was discovered with CHIME \frb{} \citep{CHIME2019rep} and its monitoring showed the first evidence for a periodicity in its activity rate of $16.35 \pm 0.18$ day \citep{CHIME2020period}.
The source was precisely localized to a nearby massive spiral galaxy at a luminosity distance of $149 \pm 0.9$\,Mpc \citep{Marcote2020Nat}, which presents an opportunity to study its host in detail and to search for possible transient/variable counterparts.
Multi-wavelength follow-up observations of \frb{} were recently performed in the radio, optical, X--ray, and gamma-ray bands \citep{Casentini2020, Panessa2020ATel, Pilia2020arXiv, Scholz2020arXiv, Tavani2020arXiv, Zampieri2020ATel}. However, no transient or variable counterpart to this source has been detected outside radio bands. 

\begin{figure}
    \centering
    \includegraphics[width=1.\columnwidth]{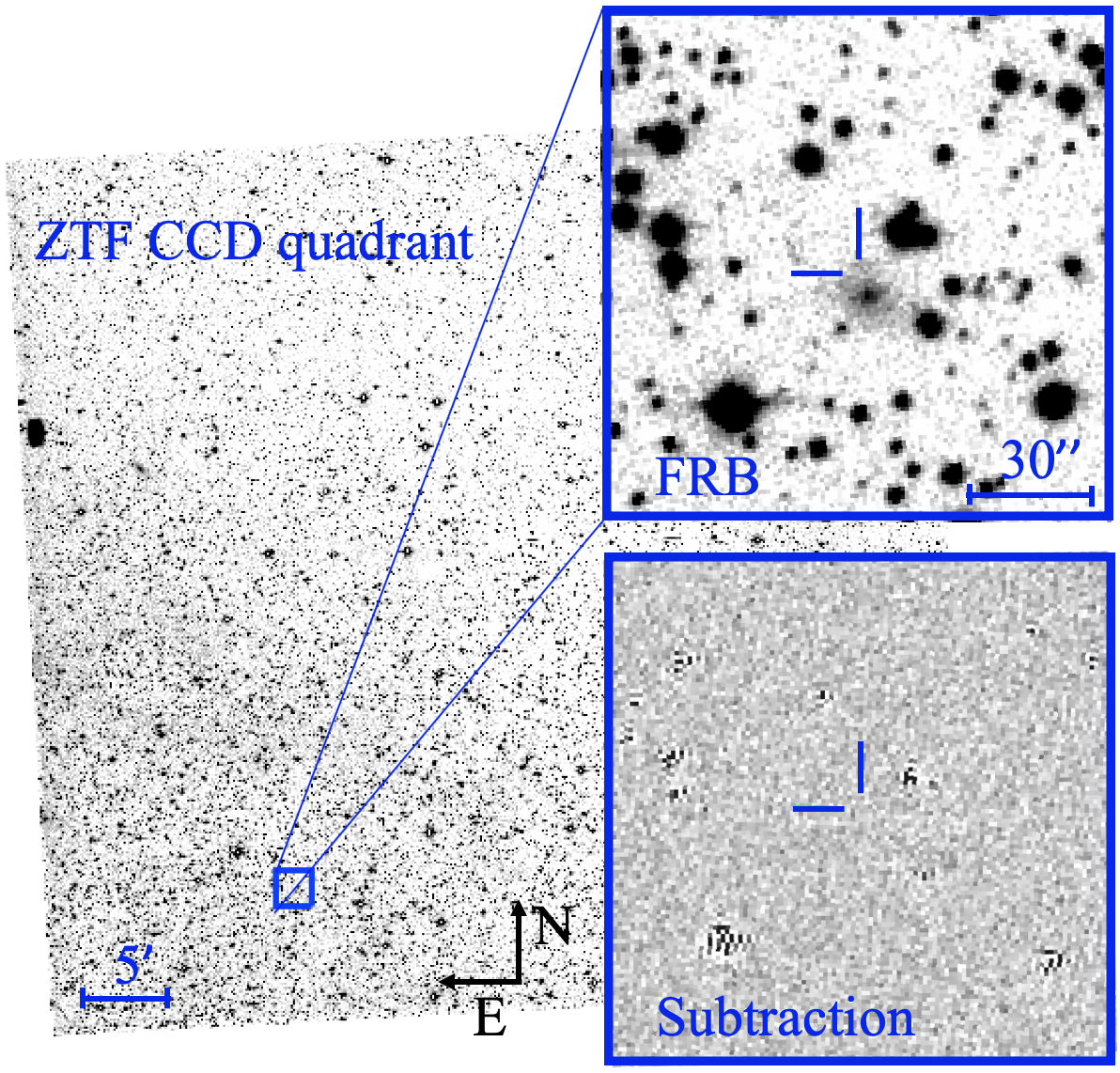}
    \caption{The location of \frb{} was observed during the ZTF survey of the Northern sky. The image shows a template image of the ZTF 3k$\times$3k pixel CCD quadrant (1 out of 64 CCD quadrants that cover 47\,deg$^2$ in total per exposure) including the FRB host galaxy. The top panel offers a closer view of the FRB location and the lower panel shows the result of image-subtraction. 
    }
    \label{fig:ZTF image}
\end{figure}

The Zwicky Transient Facility \citep[ZTF;][]{Bellm2019ZTF, Graham2019ZTF, Masci2019ZTF} on the optical Samuel Oschin 48-inch Telescope at Palomar Observatory observed the coordinates of \frb{} serendipitously during its first 2 years of activity (Figure\,\ref{fig:ZTF image}). As a result of our analysis, no variable optical counterpart to \frb{} was found in ZTF data. We describe the observations in \S\ref{sec: analysis} and the statistical method to constrain the optical-to-radio fluence ratio in \S\ref{sec: statistics}. The results are presented in \S\ref{sec: results}. We compare our constraints to previous works in \S\ref{sec: discussion} and a summary with future prospects is provided in \S\ref{sec: summary}.

\begin{figure}[t]
    \centering
        \includegraphics[width=1.\columnwidth]{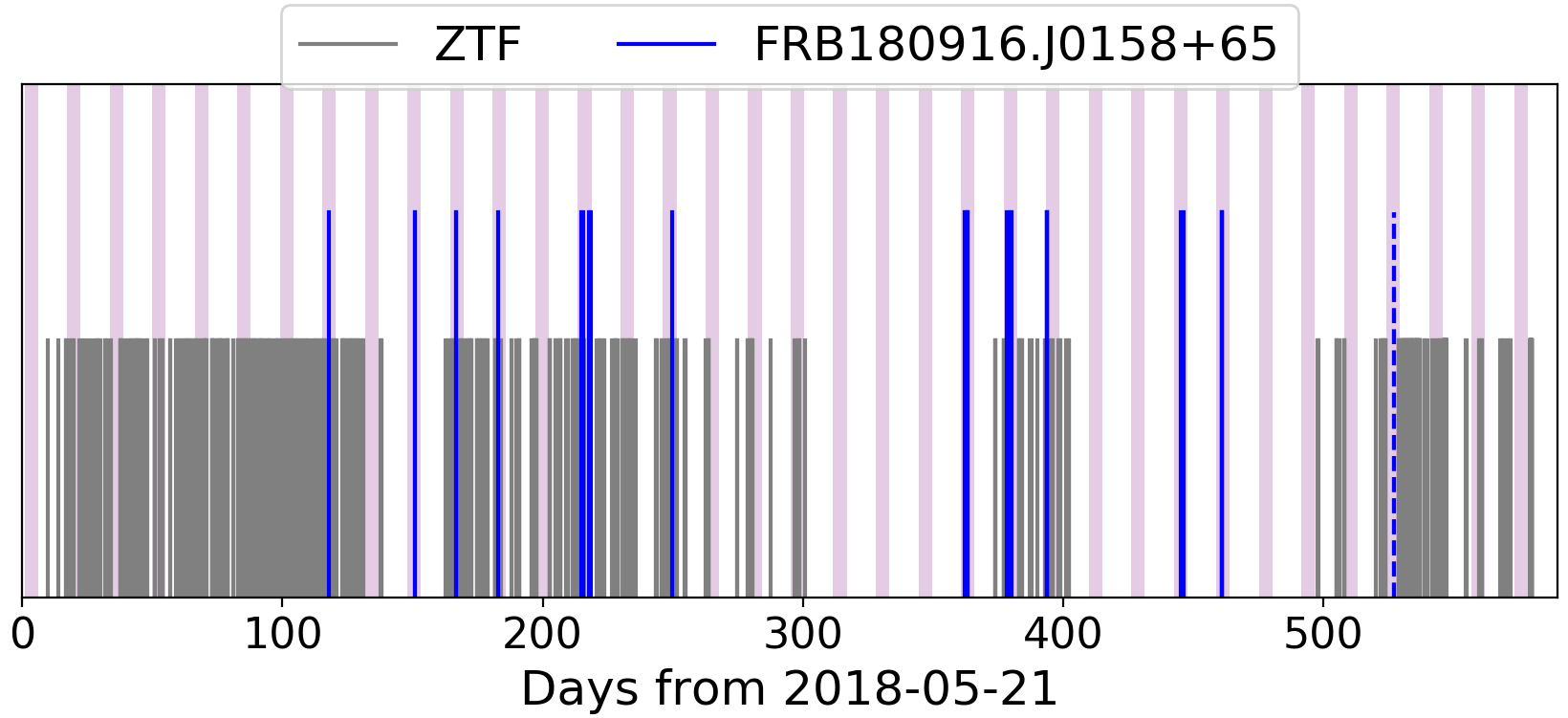}
        \includegraphics[width=0.98\columnwidth]{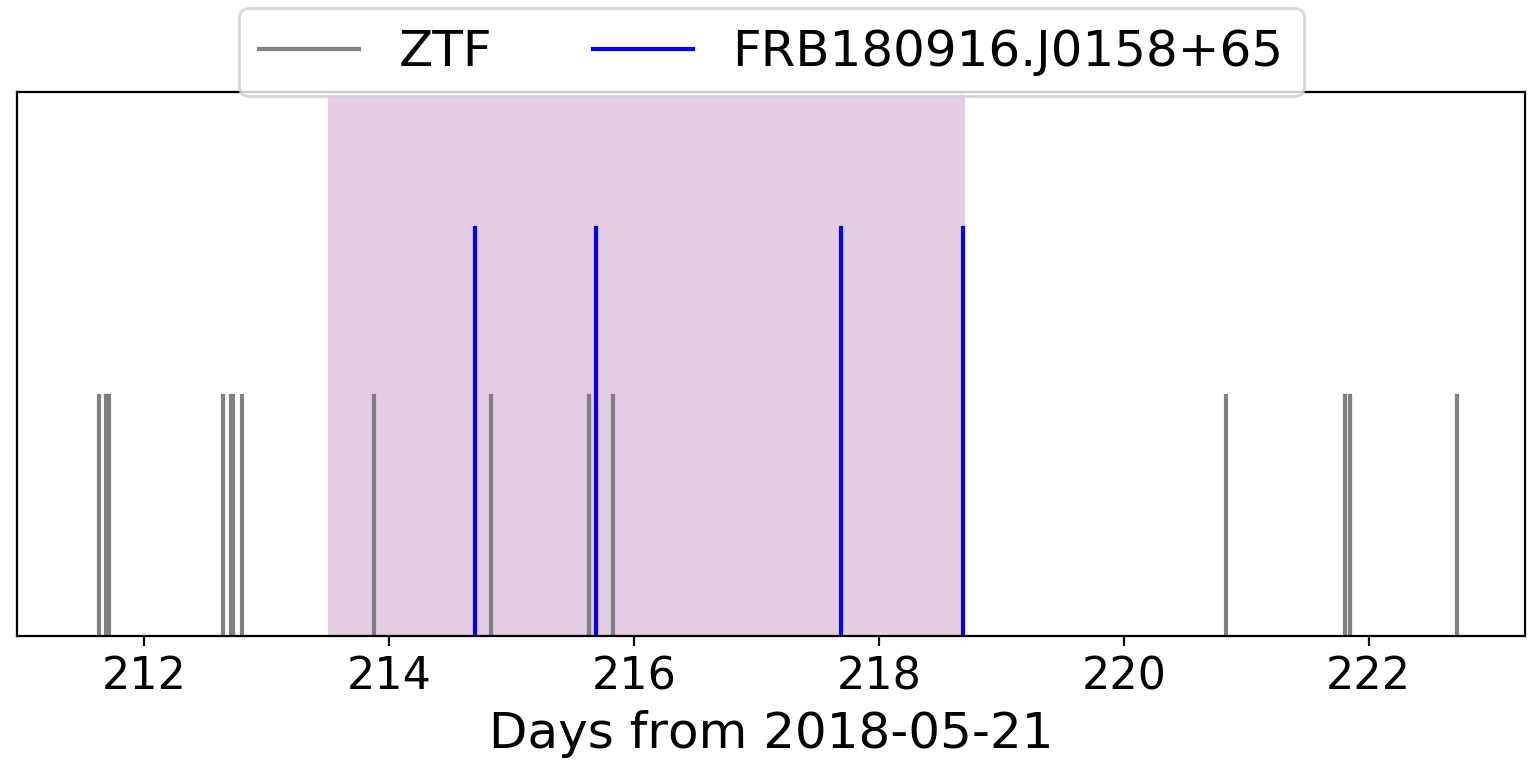}
    \caption{{\it Top --} ZTF observations (grey lines) and \frb{} detections \citep[blue lines;][]{CHIME2019rep, CHIME2020period, Marcote2020Nat}. Purple bands indicate regions of high FRB activity, determined using a period of 16.35\,d and a time window of $2\times2.6$\,d around the estimated activity peak \citep{CHIME2020period}. Dotted lines (un-resolved) indicate 2 FRBs detected on 2019 October 30, after a CHIME instrument upgrade. {\it Bottom --} Same as the {\it top} panel, zoomed in a region of the plot where 4 bursts were detected, two of which minutes to hours from ZTF observations.}
    \label{fig: observations}
\end{figure}

\section{Data analysis}
\label{sec: analysis}
ZTF surveys the sky with a typical exposure time of 30\,s per image. 
A total of 683 science images were used in this work that include the coordinates of \frb{} \citep[RAJ2000 = 01:58:00.750, DecJ2000 = 65:43:00.315;][]{Marcote2020Nat}, for a total exposure time of 5.69\,hr (Table~\ref{table: exposures}). The images were acquired from 2018 May 31 to 2019 December 22\footnote{All times in this work are expressed in \texttt{UTC}. The full observation log can be found at \url{https://www.astro.caltech.edu/~ia/log_frb_ztf.csv}} during the ZTF public survey as well as part of the ZTF partnership and Caltech surveys (P.I. Kulkarni, P.I. Prince, P.I. Graham). Figure~\ref{fig: observations} shows the temporal distribution of ZTF observations and CHIME detections of \frb{} \citep{Marcote2020Nat,CHIME2020period}.
Assuming a periodic modulation of 16.35\,d and $2\times2.6$\,d time windows centered on activity peaks that include all the FRBs published in \cite{CHIME2020period}, 147 ZTF images were acquired during high activity times of \frb{}, marked with purple stripes in Figure~\ref{fig: observations}. 
The closest ZTF exposure to an FRB detection was acquired 11.76 minutes before the FRB found on 2018 September 16 at 10:18:47.891 \citep{CHIME2020period}. Taking the frequency-dependent arrival time of the signal into account, a slightly smaller time gap of 11.69 minutes can be considered by using a dispersion measure of 349.2\,pc\,cm$^{-3}$ \citep{CHIME2019rep}, which would cause a burst to be detected in the optical bands 4.025\,s before reaching the center of the CHIME band at 600\,MHz, i.e. the reference frequency for the reported burst times.

\begin{table*}
    \centering
    \begin{tabular}{cccccccccccc}
    \hline \hline
    images & filter & exptime & mag & mag corrected & seeing & $F_{50}$ &  $F_{90}$ &  $F_{95}$ & $\eta_{50}$ & $\eta_{90}$ & $\eta_{95}$\\
     & & (s) & (AB) & (AB) & & (Jy\,ms) & (Jy\,ms) & (Jy\,ms) & & & \\
    \hline
    227 & $g$ & 6810 & 20.72 & 17.39 & all & 12.11 & 31.17 & 39.97 & 1.2 & 3.1 & 4.0\\
    173 & $g$ & 5190 & 20.88 & 17.56 & $< 3''$ & 10.55 & 27.87 & 37.488 & 1.1 & 2.8 & 3.7 \\
    456 & $r$ & 13680 & 20.22 & 17.92 & all  & 8.02 & 201.11 & 307.12 & 0.8 & 20.1 & 30.7 \\
    363 & $r$ & 10890 & 20.41 & 18.11 & $< 3''$ & 7.33 & 58.49 & 110.90 & 0.7 & 5.8 & 11.1\\
    683 & $g+r$& 20490 & - & - & all & 9.92 & 133.78 & 246.26 & 1.0 & 13.4 & 24.6\\
    536 & $g+r$& 16080 & - & - & $< 3''$ & 8.25 & 42.06 & 96.99 &  0.8 & 4.2 & 9.7 \\   
    \hline
    \end{tabular}
    \caption{The first 5 columns indicate the number of 30\,s images analyzed, the filter, the total exposure time, median limiting magnitude (50\% confidence) before and after Galactic extinction correction \citep{Schlafly2011}. The quantities are calculated using images taken under all seeing conditions and using only those with seeing $<3''$. Then we report the fluence limit at 50\%, 90\%, and 95\% confidence. Finally, we indicate the ratio between optical and radio fluence assuming an FRB fluence $\mc{F}_{\rm frb} = 10$\,Jy\,ms.}
    \label{table: exposures}
\end{table*}

The precise localization of the FRB made it possible to perform forced point-spread function (PSF) photometry in all the available images, assuming that an optical burst would have indeed a PSF profile. A limit based on forced photometry is more reliable than a limit based on the lack of ZTF ``alerts" \citep{Bellm2019ZTF} because alerts are affected by blind detection efficiency, which by definition is $\sim 50\%$ around the limiting magnitude, and because forced photometry allows us to lower the discovery threshold. We used \texttt{ForcePhot} \citep{Yao2019} to perform forced PSF photometry on images processed with the ZTF real-time reduction and image subtraction pipeline at the Infrared Processing \& Analysis Center \citep{Masci2019ZTF} using the \texttt{ZOGY} algorithm \citep{Zackay2016}. Image subtraction helped us obtain cleaner photometry by removing host galaxy flux and reducing stars crowdedness in the field (Figure\,\ref{fig:ZTF image}). 
Upon non-detection of a source at 3$\sigma$ level, we considered conservative upper limits calculated by the ZTF pipeline.

\begin{figure}[t]
    \centering
        \includegraphics[width=1.\columnwidth]{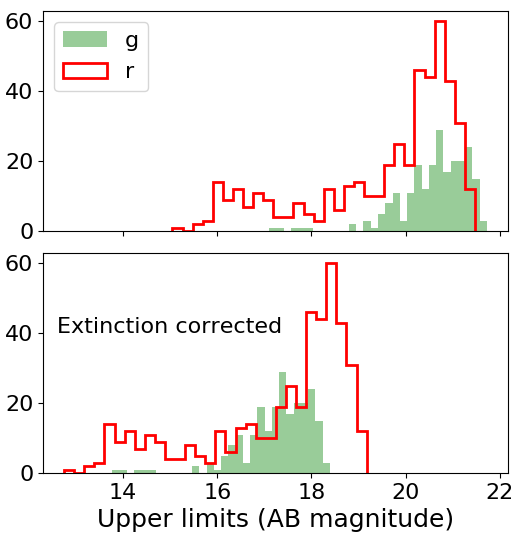}
    \caption{{\it Top --} Distribution of $3\sigma$ upper limits (apparent magnitude in AB system) on 683 ZTF images. {\it Bottom --} the same distribution is shown after correcting for Galactic extinction along the line of sight using the \citep{Schlafly2011} dust maps. }
    \label{fig: mag limit}
\end{figure}

Forced photometry on 683 images returned 3 faint detections (on 2018 July 22 11:00:27, 2018 September 28 07:36:45, and 2018 December 31 02:26:04) and 680 non-detections. The three $\sim 3\sigma$ detections were 
deemed spurious because of imperfect image subtraction across the chip and the shape of the residuals hardly approximating a PSF. In conclusion, no optical burst from \frb{} was found in ZTF images. We also inspected the images to check that no sign of saturation (which may result in unreliable photometry) was present and that no cosmic ray (CR) hit the CCD at the FRB location during ZTF exposures, since the charge deposition of a CR could mimic the occurrence of an optical flash. With a hit rate of 2\,cm$^{-2}$\,min$^{-1}$, we could expect about 21 CRs per $3072\times3072$ pixels quadrant per 30\,s frame. We corrected the apparent magnitudes for the Galactic dust extinction along the line of sight using the \cite{Schlafly2011} dust maps. \frb{} is located at low Galactic declination ($b=3.72$), which explains the large extinction of $E(B-V) = 0.87$ \citep{Schlafly2011}. We note that a lower extinction of $E(B-V) = 0.68$ is obtained using the 3D dust map based on {\it Gaia}, Pan-STARRS 1 and 2MASS \citep{Green2019dust}, so our results obtained using the \cite{Schlafly2011} extinction are conservative.
The distribution of the 3$\sigma$ upper limits that we obtained is shown in Figure~\ref{fig: mag limit}.

\section{Statistical analysis}
\label{sec: statistics}
The objective of the proposed method is to yield statistically robust constraints, under a few assumptions, despite the fact that all detected repetitions of \frb{} occurred outside ZTF exposures. In particular, we assume that every FRB is accompanied by a short-duration ($\ll 30\,$s) optical flash with fluence $\eta$ times the FRB fluence, then we calculate the limiting probability $\t{P}(\eta)$ as given by ZTF non-detection of optical flashes for each $\eta$. Radio (FRB) fluence is denoted as $\mc{F}$, so the corresponding optical fluence is $F=\eta\mc{F}$. The results obtained in this section are agnostic to any assumption of periodic modulation, hence they are insensitive to possible period aliasing.
The detailed procedure is as follows.

We start from the cumulative fluence distribution $N_{\rm frb}(\geq \mc{F})$ of the CHIME bursts as shown in the Extended Data Figure 3 of \citet{CHIME2020period}. A fluence threshold $\mc{F}$ is chosen such that there are $N_{\rm frb}$ detections above it. The goal is to construct a statistical model which reproduces the $N_{\rm frb}$ CHIME bursts while determining the number of accompanying optical flashes captured by our ZTF images at the same time. We assume that the time intervals $\delta$ between adjacent bursts (above a certain fluence) follows the Weibull distribution, which has been used to model the first repeater FRB 121102 \citep{Oppermann2018} and the whole CHIME repeater population \citep{Lu2020}. The cumulative distribution of wait time $\delta$ is given by
\begin{equation}
    P(<\delta)  = 1 - \mr{exp}\left[-(r\delta\, \Gamma(1+1/k))^k\right],
\end{equation}
where $k$ is the Weibull clustering parameter ($k<1$ describes that small intervals are favored compared to the Poissonian, $k=1$, case), $r=\lara{\delta}^{-1}$ is the mean repeating rate, and $\Gamma(x)$ is the gamma function. Note that $k\simeq 1/3$ is favored by analysis of the first repeater FRB 121102 \citep{Oppermann2018, Oostrum20}.

Our $t=0$ corresponds to 2018-05-31 11:19:32 when the first ZTF exposure starts, and then the starting time of the $l$-th ZTF observation is denoted $t^{\rm ztf}_l$ (with $t^{\rm ztf}_0=0$ by definition and $l=0$, 1, \ldots, 682). We randomly draw successive time intervals $\delta_i$ such that the $n$-th FRB and optical flash occur at time $t_n = t_0 + \sum_{i=0}^{n}\delta_i$, where $t_0$ is a negative random number between $-30/r$ and $-10/r$ such that the first ZTF exposure time ($t=0$) is far from the start of burst series ($t=t_0$). CHIME exposures start on 2018 August 28 \citep{CHIME2020period} and occur regularly on a (sidereal) daily basis, so we take the starting time for each CHIME observation to be $t^{\rm chime}_m= (90.5+\xi m)\,$d (the precise value of $t^{\rm chime}_0$ is unimportant), where $\xi\equiv 23.9344696/24$ is the scale factor for the length of a sidereal day. With three time series \{$t_n$\} (FRB occurrence, randomly generated), \{$t^{\rm ztf}_l$\} (start time of ZTF exposures, fixed), and \{$t^{\rm chime}_m$\} (start time of CHIME daily on-sky exposures, fixed), the $n$-th burst is recorded as detectable by ZTF (or CHIME) if $t^{\rm ztf}_l<t_n< t^{\rm ztf}_l + \Delta^{\rm ztf}$ (or $t^{\rm chime}_m<t_n< t^{\rm chime}_m + \Delta^{\rm chime}$), where $\Delta^{\rm ztf}=30\,$s and $\Delta^{\rm chime}=12\,$min are the durations of each ZTF and CHIME daily exposures, respectively. We use the CHIME data up to 2019 September 30, so the index for CHIME is $m=0$, 1, \ldots, 398. During this time, observations were sometimes interrupted by testing, so we randomly take away a fraction $f_{\rm off}=0.16$ of the daily exposures, meaning that no detection is recorded for the off-line day even if $t^{\rm chime}_m<t_n< t^{\rm chime}_m + \Delta^{\rm chime}$. This reduction gives total exposure time is of 2.7 days or 64 hours consistent with the estimate by \citet{CHIME2020period}. We exclude two CHIME bursts detected on 2019 October 30 (dashed blue lines in Figure \ref{fig: observations}) in our repeating rate estimation, because they are detected after a major pipeline upgrade.

With the above procedure, for each set of parameters \{$r$, $k$\} (mean repeating rate and temporal clustering), we carry out $N_{\rm cases}=1000$ random cases and determine  the detected number of bursts by CHIME ($N_{\rm chime}$) and the number of optical flashes within the ZTF coverage ($N_{\rm ztf}$). The likelihood that this set of parameters reproduces the CHIME data is given by $L=\sum(N_{\rm chime}=N_{\rm frb})/N_{\rm cases}$, where $\sum(N_{\rm chime}=N_{\rm frb})$ is the number of cases that match the number of observed bursts above a certain fluence $\mc{F}$. We have tested that $N_{\rm cases}$ is sufficiently large to yield a stable likelihood whose random fluctuation is negligible compared to other uncertainties of our problem. More realistically, the likelihood should also include the goodness of fit between the simulated and observed distributions of time intervals between adjacent bursts, which will constrain the Weibull parameter $k$ \citep{Oppermann2018, Oostrum20}. Instead, we take a number of different fixed $k\in (1/4, 1/2)$ and treat the resulting difference as the systematic error of our method. This is motivated by the fact that the expected number of ZTF detections mainly depends on the mean repeating rate $r$. For the same reason, the 16-day periodicity of \frb{} raises the mean repeating rate within the $\pm$2.6-day (the exact numbers are unimportant) active window by a factor of 16/5.2 in \textit{both} the CHIME and ZTF observing runs coincident with the active windows, so our method is insensitive to the periodicity. Then, for each $k$, we use the following Markov Chain Monte Carlo (MCMC) method to constrain log$\,r$[d$^{-1}$] and simultaneously determine the probability $P(N_{\rm ztf}\geq 1)$. Hereafter the mean repeating rate $r$ is in units of d$^{-1}$.

The initial value is taken to be log$\,\bar{r}$ and we assume a flat prior of log$\,r\in (\mr{log}\bar{r}-1, \mr{log\,}\bar{r}+1)$, where $\bar{r}=N_{\rm frb}/2.7\,$d is the mean expectation.
We record the probability of at least one optical flash occurring within ZTF exposures $P_i(N_{\mr{ztf}, i}\geq 1)$ for each accepted sample of $\mr{log}\,r_i$. Finally, for an assumed fluence ratio $\eta$ between optical flashes and FRBs, the probability that ZTF captured at least one optical flashes brighter than $\eta \mc{F}$ is given by
\begin{equation}
    P_{\rm d}(\mc{F}, \eta) = \sum_i P(N_{\mr{ztf}, i}\geq 1)/N_{\rm samp},
\end{equation}
where $N_{\rm samp}$ is the total number of accepted MCMC samples (convergence is achieved for $N_{\rm samp}\gtrsim 3\times10^5$). The above probability is a decreasing function of the FRB fluence cut $\mc{F}$, because $N_{\rm frb}(\geq \mc{F})$ and hence the mean repeating rate $r$ decrease with $\mc{F}$.

On the other hand, our non-detection in all 683 ZTF images rule out any flash brighter than $\eta \mc{F}$ (beyond 3$\sigma$) at probability $P_{\rm nd}(\eta \mc{F})$. Thus, we are able to rule out the particular fluence ratio $\eta$ with survival probability $\t{P} =1- P_{\rm d}(\mc{F}, \eta)P_{\rm nd}(\eta \mc{F})$. For a given $\eta$, we try a number of different fluence cuts $\mc{F}\in (6, 20)\,\rm Jy\, ms$ and the corresponding observed $N_{\rm frb}(\geq \mc{F})$, and the best constrained case gives the lowest survival probability $\t{P}(\eta)$. Below the CHIME completeness threshold $\mc{F}\simeq 6\,\rm Jy\,ms$, instead of taking the detected number, we use power-law $N_{\rm frb}(\geq \mc{F}) \propto \mc{F}^{1-\gamma}$ extrapolation. Steeper power-law (larger $\gamma$) will give more FRBs at low fluences and hence stronger constraints on $\eta$. We show the results for $\gamma=1.8$, as motivated by the study of the CHIME repeating sample \citep{Lu2020}, and for $\gamma=2.4$, as indicated by the apparent fluence distribution of \frb{}  \citep{CHIME2020period}. 

\begin{figure}
    \centering
    \includegraphics[width=1.\columnwidth]{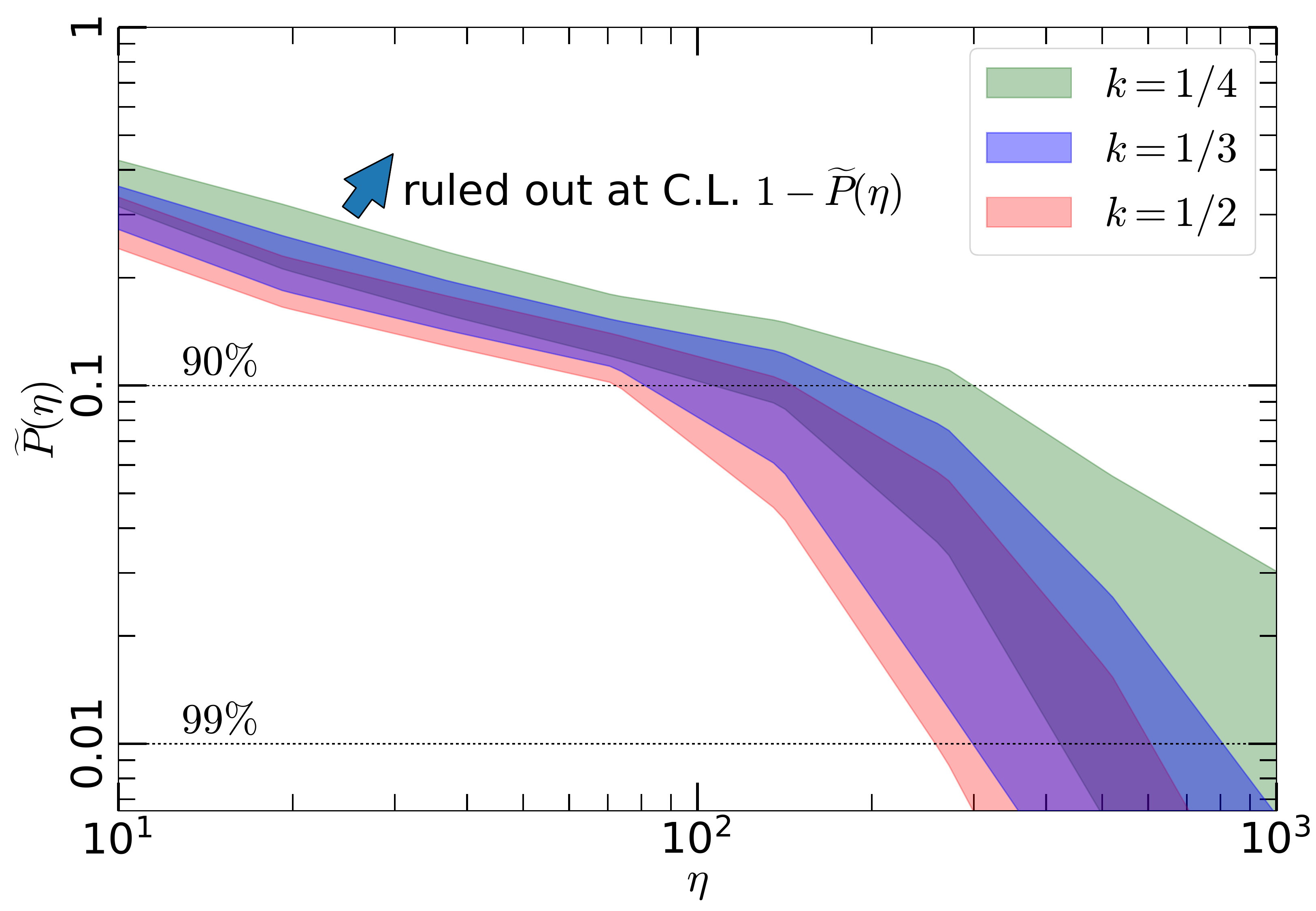}
    \caption{Limits on the optical-to-radio fluence ratio $\eta=F/\mc{F}$ from non-detection of optical flashes with ZTF. The probability that $\eta$ is allowed is denoted as $\t{P}(\eta)$, so each $\eta$ is ruled out at confidence level (C.L.) of $1-\t{P}(\eta)$. The three colored bands are for different $k=1/4$ (green),$1/3$ (blue, our fiducial case), $1/2$ (red), with decreasing $\t{P}(\eta)$. The upper and lower bounds for each band are for $\gamma=1.8$ and $2.4$, respectively. Generally, we rule out $\eta\simeq 200$, or $E_{\rm opt}\simeq 3\times 10^{46}\,$erg associated with an FRB with $\mc{F}=10\rm\,Jy\,ms$, at 90\% confidence level.
    }
    \label{fig:Peta}
\end{figure}

\section{Results}
\label{sec: results}

The analysis described in \S\ref{sec: statistics} provides a framework to robustly combine ZTF measurements and CHIME detections of a repeating FRB, even if the detected radio bursts were not coincident with ZTF observations of the source. As shown in Fig. \ref{fig:Peta}, we are able to constrain $\eta \lesssim 200$ at 90\% confidence level, corresponding to a limit on the energy of $E_{\rm opt} \lesssim3\times 10^{46}\,$erg for an optical counterpart associated with a $\mc{F}=10\rm\,Jy\,ms$ FRB.

The results can be understood as follows. The expectation value of the repeating rate above threshold $F_{\rm th} = 5\,\rm Jy\, ms$ is about $1\rm\, hr^{-1}$ within the $\pm$2.6-day active windows \citep{CHIME2020period}. If we simply \textit{assume} a rate of $r(>5\,\rm Jy\, ms)=1\rm\, hr^{-1}$, then on average $1.2\pm1.1$ optical flashes are expected to occur in our 147 ZTF images acquired within the active windows. The ZTF limits in those time frames yield $F_{95}\simeq 30 \rm \,Jy\, ms$ (95\% confidence), which means that we can rule out $\eta = F_{95}/\mc{F}_{\rm th}\simeq 3$ at 95\% confidence. However, the repeating rate of $r(>5\mr{\,Jy\,ms})$ is highly uncertain. 
Our method in \S\ref{sec: statistics} combines the probability distributions of both the repeating rate and the ZTF limiting fluence and hence gives a robust statistical constraint on $\eta$ that is a factor of $\sim100$ less stringent than the above simple expectation.

\section{Discussion}
\label{sec: discussion}

Optical observations of \frb{} with high cadence were performed by \cite{Zampieri2020ATel} using the fast optical photon counter IFI+IQUEYE \citep{Naletto2009} mounted on the 1.2-m Galileo telescope. Although no FRB was detected during their observations, \cite{Zampieri2020ATel} placed an upper limit on the optical fluence of $F \lesssim 0.151$\,Jy\,ms\footnote{\label{note: galactic} Galactic extinction correction is not addressed.}, which is more constraining than ZTF results thanks to the significantly shorter exposure time. \cite{Zampieri2020ATel} also obtained results$^{\ref{note: galactic}}$ comparable with ZTF using the 67/92 Schmidt telescope near Asiago, Italy. 

\cite{Hardy2017} place constraints on the optical fluence of FRB~121102 using 70\,ms exposures coincident with radio observations, obtaining $E_{\rm opt} \lesssim 10^{43}$\,erg (or equivalent fluence ratio $\eta\lesssim 0.02$ for the brightest FRB). Similarly, \cite{MAGIC2018} observed FRB~121102 using the Major Atmospheric Gamma
Imaging Cherenkov simultaneously with Arecibo. Along with 5 radio bursts, they detected no optical $U$ band bursts with fluence $> 9 \times 10^{-3}$\,Jy\,ms$^{\ref{note: galactic}}$ at 1\,ms exposure time bins, although 1 optical burst was found 4\,s before an FRB, with 1.5\% random association probability.
We note that there were no explicit mention to Galactic extinction correction in previous work discussed in this section. FRB~121102 is also located close to the Galactic plane ($b = -0.22$\,deg), where the extinction is significant. The effect should be smaller for \cite{Hardy2017} than for ZTF $g$- and $r$-band observations, thanks to their redder $i'+z'$ broadband filter. The excellent limit placed by \cite{MAGIC2018} may suffer a larger correction for the $U$ band, which should be taken into account in multi-wavelength FRB modeling. 

Our current 90\% limit of $\eta\simeq 200$ corresponds to an optical-to-radio \textit{energy} ratio of $\eta \nu_{\rm opt}/\nu_{\rm frb}\sim 10^8$, which is at least two to three orders of magnitude above the predictions of the synchrotron maser model based on magnetar flares \citep[assuming pair-dominated upstream plasma;][]{Metzger19, Beloborodov19}. However, our constraint can be significantly improved with shorter per-image exposure time and longer total integration time in future observations. We also note that, assuming the same observing sequence, ZTF observations at high Galactic latitude (with negligible dust extinction) would yield an order of magnitude deeper constraints on the fluence.

\section{Summary and future prospects}
\label{sec: summary}

In this work, we placed constraints on the optical fluence of \frb{} using 683 images serendipitously acquired with ZTF. The statistical analysis presented in \S\ref{sec: statistics} combined disjoint ZTF exposures and CHIME detections into a robust upper limit on optical-to-radio fluence ratio  $\eta \lesssim 200$ (90\% confidence) and on the emitted energy in the optical $E_{\rm opt} \lesssim 3\times 10^{46}$\,erg for a 10~Jy~ms FRB.

This work further demonstrated that possible optical counterparts to FRBs can be constrained with large field-of-view optical surveys such as ZTF, TESS \citep[see also the search for optical counterparts to FRB~181228 by][]{Tingay2019}, and soon the Legacy Survey of Space and Time (LSST) at Vera Rubin Observatory. As Table\,\ref{table: exposures} suggests, ZTF has the potential of placing deeper constraints when bright ($\mc{F} > 10$\,Jy\,ms) FRBs are simultaneously observed, especially at high Galactic latitudes. 
New high-cadence instruments can also play an important role in FRB counterpart detection with sub-second observations. These include drift scan imaging experiments \citep{Tingay2020PASA}, the wide-field Tomo-e Gozen instrument \citep{Sako2016, Richmond2020} or the Weizmann Fast Astronomical Survey Telescope \citep[WFAST;][]{Nir2017AAS} based on complementary metal-oxide-semiconductor (CMOS) technology, or the Ultra-Fast Astronomy \citep[UFA;][]{Li2019} observatory that will observe variable sources at millisecond to nanosecond timescales using two single-photon resolution fast-response detectors.

This work also showed that nearby, highly active FRBs such as \frb{} present us with ground for statistical estimate of optical fluence limits even without simultaneous optical+radio observations. Dedicated high-cadence experiments may have higher chances of detecting optical flashes from FRBs, for example using the Caltech HIgh-speed Multi-color camERA \citep[CHIMERA;][]{Harding2016} mounted at the prime focus of the large 200-inch Hale Telescope at Palomar Observatory. CHIMERA could yield constraints up to $100$ times deeper than any existing optical observation to date. On the other hand, \cite{Chen2020arXiv} suggest a different method to quantify the presence of FRB counterparts in the existent datasets of large, continuous multi-wavelength and multi-messenger surveys. In conclusion, the near future offers a plenty of opportunities to further investigate optical counterparts to FRBs.

\acknowledgements
\section*{Acknowledgments}

We thank the anonymous referee for their comments that improved the quality of the paper. We thank Shri Kulkarni, Sterl Phinney, Gregg Hallinan, Vikram Ravi, Dana Simard, Jesper Sollerman, and Eran Ofek for fruitful discussion.
I.~A. thanks Kendrick Smith for his inspiring colloquium at Caltech, and Benito Marcote and Wael Farah for useful communication.
Based on observations obtained with the Samuel Oschin 48-inch Telescope and the 60-inch Telescope at the Palomar Observatory as part of the Zwicky Transient Facility project, a  scientific  collaboration  among  the  California  Institute  of Technology,  the  Oskar  Klein  Centre,  the  Weizmann  Institute of Science, the University of Maryland, the University of Washington, Deutsches Elektronen-Synchrotron, the University of Wisconsin-Milwaukee, and the TANGO Program of the University System of Taiwan.  Further support is provided by the U.S. National Science Foundation under Grant No. AST-1440341. The ZTF forced-photometry service was funded under the Heising-Simons Foundation grant \#12540303 (PI: Graham).
The ztfquery code was funded by the European Research Council (ERC) under the European Union's Horizon 2020 research and innovation programme (grant agreement n759194 - USNAC, PI: Rigault).
This work was supported by the GROWTH (Global Relay of Observatories Watching Transients Happen) project funded by the National Science Foundation under PIRE Grant No 1545949. GROWTH is a collaborative project among California Institute of Technology (USA), University of Maryland College Park (USA), University of Wisconsin Milwaukee (USA), Texas Tech University (USA), San Diego State University (USA), University of Washington (USA), Los Alamos National Laboratory (USA), Tokyo Institute of Technology (Japan), National Central University (Taiwan), Indian Institute of Astrophysics (India), Indian Institute of Technology Bombay (India), Weizmann Institute of Science (Israel), The Oskar Klein Centre at Stockholm University (Sweden), Humboldt University (Germany), Liverpool John Moores University (UK) and University of Sydney (Australia).

\begin{appendix}
\section{Optical flash detectability with ZTF}
\label{subsec: optical flash detectability}
In this section, we describe the response of ZTF CCDs if bright optical flashes are observed.
The \emph{creation} of electron hole pairs when photons are absorbed is effectively instantaneous and is linear, i.e. the absorption probability is not affected by the photon flux since the density of valence band electrons available to be promoted to the conduction band is very high compared to the photon flux.
The \emph{collection} of photo-generated charge is lossless and thus linear only if clock voltages are set to prevent charge from interacting with traps at the surface.  For present clock settings in ZTF, this is not the case during exposure since clocks are positively biased to minimize lateral charge diffusion at the expense of linearity beyond saturation. However, we visually checked all the processed images and none showed signatures of saturation. 

A possible effect of very high flux is electrostatic repulsion of photo-generated charges.  An illustrative example of this effect occurs when (rare) $\alpha$ particles generated by radioactive decay of Uranium or Thorium occur in the bulk silicon \citep{Aguilar-Arevalo2015arXiv}. Having energies $\sim4$\,MeV, these $\alpha$ particles deposit $\gtrsim4$ million electrons, since bandgap in silicon is 1.14\,eV. These $\alpha$ particle events cover a circular area whose size depends on the thickness of the field free region.  The effect is well documented in thick fully depleted CCDs and typically spans 5-6 pixels \citep{Aguilar-Arevalo2015arXiv}. 

Data for the equivalent electrostatic repulsion prior to charge collection is scarce in the standard CCDs used in ZTF.  These have a field free region near the back surface which is estimated to vary from 10 to 20 $\mu m$ in a radial pattern, based on overall thickness implied by surface metrology and confirmed by fringing patterns from night sky lines. 
Once the charge diffuses towards the potential wells, the PSF depends on signal in well-known ways: brighter-fatter effect \citep{Guyonnet2015}, charge blooming, tails due to charge transfer inefficiency, and signal non-linearity. These effects are not expected to significantly affect the limits calculated in this work.
\end{appendix}


\end{document}